\documentclass[twocolumn,showpacs,preprintnumbers,amsmath,amssymb]{revtex4}
\usepackage{graphicx}% Include figure files
\usepackage{dcolumn}% Align table columns on decimal point
\usepackage{bm}% bold math

\begin{document}

%\preprint{APS/123-QED}

\title{Novel vortex structures in dipolar condensates}
\author{S. Yi and H. Pu}

\affiliation{Department of Physics and Astronomy, and Rice Quantum
Institute, Rice University, Houston, TX 77251-1892, USA}

\begin{abstract}
We investigate the properties of single vortices and of vortex
lattice in a rotating dipolar condensate. We show that vortices in
this system possess many novel features induced by the long-range
anisotropic dipolar interaction between particles. For example,
when the dipoles are polarized along the rotation axis, vortices
may display a crater-like structure; when dipoles are polarized
orthogonal to the rotation axis, vortex cores takes an elliptical
shape and the vortex lattice no longer possesses hexagonal
symmetry.
\end{abstract}

\date{\today}
\pacs{03.75.Lm, 03.75.Nt}

\maketitle

Quantized vortices represent one of the fundamental topological
structures in nature. Studies of quantized vortices were launched
out of the efforts to understand the flow properties of superfluid
$^4$He, but have now extended into a variety of systems including
superconductors, superfluid $^3$He, quantum magnets, liquid
crystals, nuclear matter and even cosmic strings \cite{vorall}.
The realization of weakly-interacting quantum gases of atomic
vapors provides a new platform for the study of quantum vortices.
Vortices in atomic Bose-Einstein condensates have been produced,
perturbed and engineered by precise external fields, and can be
studied theoretically in a quantitative manner using
first-principles calculations \cite{fetter}. Very recently,
vortices were also observed in a quantum degenerate Fermi gas
\cite{vorfermi}.

So far, vortices have only been produced in condensates featuring
short-range $s$-wave collisions. The creation of condensation in
the transition metal $^{52}$Cr offers a completely new system of
quantum gases \cite{crBEC}. The ground state of $^{52}$Cr has a
large magnetic dipole moment of $\mu = 6\mu_B$, with $\mu_B$ being
the Bohr magneton. Hence the magnetic dipolar interaction between
atoms plays a crucial role in $^{52}$Cr condensates. Previous
theoretical studies of dipolar quantum gases have focused on the
ground state properties of non-rotating systems \cite{dipreview}.
The purpose of this paper is to investigate the properties of
vortices in a rotating dipolar condensate. As we will show, due to
the long-range and anisotropic nature of the dipolar interaction,
single vortices and vortex lattices in such systems possess many
unique features not present in condensates with only contact
collisions.

{\em Model} --- We consider a trapped condensate of $N$ $^{52}$Cr
atoms whose dipole moments are polarized along certain direction.
The two-body dipolar interaction potential is
\begin{eqnarray}
V_{dd}({\mathbf r},{\mathbf
r}')=\frac{\mu_0\mu^2}{4\pi}\,\frac{1-3\cos^2\theta}{|{\mathbf
r}-{\mathbf r}'|^3}, \label{vdd3d}
\end{eqnarray}
where $\mu_0$ is the vacuum magnetic permeability, and $\theta$ is
the angle between the dipole moment and the vector ${\mathbf
r}-{\mathbf r}'$. The trapping potential is assume to be harmonic
with axial symmetry
\begin{eqnarray}
U({\mathbf r})=\frac{1}{2}M\omega_\perp^2(x^2+y^2+\lambda^2z^2)
\end{eqnarray}
with $\omega_\perp$ being the radial trap frequency and $\lambda =
\omega_z /\omega_\perp$ the trap aspect ratio. For the sake of
simplicity, we will focus on a pancake-shaped system with $\lambda
\gg 1$ such that the condensate can be regarded as quasi
two-dimensional (2D) whose motion along the $z$-axis is frozen to
the ground state of the axial harmonic oscillator. Therefore the
condensate wave function can be decomposed as $\Psi({\mathbf
r},t)=\psi({\boldsymbol\rho},t)\phi(z)$, where
${\boldsymbol\rho}=(x,y)$ and
$\phi(z)=(\lambda/\pi)^{1/4}\,e^{-\lambda z^2/2}$. Throughout this
paper, we adopt a unit system where the units for length, frequency
and energy are given by $a_\perp=\sqrt{\hbar/M\omega_\perp}$,
$\omega_\perp$ and $\hbar\omega_\perp$, respectively. The reduction
of the effective spatial dimensionality considerably reduces the
computational intensity. Although intrinsic three-dimensional (3D)
phenomena such as the bending and Kelvin mode excitation of vortex
lines cannot be studied, the 2D model does capture many essential
features of vortices at equilibrium, for example, the structure of
the vortex core and that of vortex lattice.

We assume that the condensate is rotating along the symmetry axis
$z$ with angular frequency $\Omega$. After integrating out the
$z$-variable, we obtain the effective 2D Gross-Pitaevskii equation
in the rotating frame as
\begin{eqnarray}
i\frac{\partial \psi}{\partial
t}&=&\Big[-\frac{\nabla_\perp^2}{2}+\frac{\rho^2}{2}-\mu-\Omega L_z
+g_{2D}|\psi|^2\nonumber\\
&&+\int d{\boldsymbol\rho}'\,|\psi({\boldsymbol\rho}')|^2\,
V_{dd}^{2D}({\boldsymbol\rho},{\boldsymbol\rho}')
\Big]\psi({\boldsymbol\rho}),\label{gp2d}
\end{eqnarray}
where $\mu$ is the chemical potential, $L_z$ the $z$-component of
the orbital angular momentum operator,
$g_{2D}=2(2\pi\lambda)^{1/2}Na/a_\perp$ the rescaled 2D
collisional interaction strength with $a$ being the $s$-wave
scattering length which can be tuned via Feshbach resonance
\cite{crFesh}. The rescaled 2D dipolar interaction potential takes
the form
\begin{equation}
\label{dd2d} V_{dd}^{2D}({\boldsymbol\rho},{\boldsymbol\rho}') =
\int dz \,|\phi(z)|^4\, V_{dd}({\bf r}, {\bf r}').
\end{equation}
In the following, we will treat separately the cases where the
dipoles are polarized along the axial and along the transverse
direction, as they lead to very different phenomena.

\begin{figure}
\centering
\includegraphics[width=2.7in]{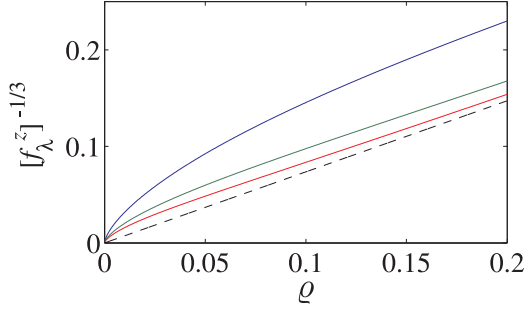}
\caption{(Color online). Solid lines: function $
[f^z_{\lambda}(\varrho)]^{-1/3}$ for $\lambda=11, 100$, and $400$ in
descending order. The dashed line represents the asymptote
$(2\pi)^{-1/6} \varrho$.} \label{vdip2dz}
\end{figure}

{\em Axially polarized dipoles} --- For this case, we can find
from Eq.~(\ref{dd2d}) that
\begin{eqnarray}
V_{dd}^{2D}({\boldsymbol\rho},{\boldsymbol\rho}')
=\frac{N\mu_0\mu^2}{(2\pi)^{3/2}a_\perp^3\hbar\omega_\perp}\,f^z_\lambda(\varrho)\,,
\label{dip2d}
\end{eqnarray}
where $\varrho=|{\boldsymbol\rho}-{\boldsymbol\rho}'|$ and
\begin{eqnarray}
f^z_\lambda(\varrho)&=&\frac{\lambda^{3/2}e^{\lambda\varrho^2/4}}{6}
\left[\lambda\varrho^2K_0
\left(\frac{\lambda\varrho^2}{4}\right)+\left(2-\lambda\varrho^2\right)\right.\nonumber\\
&&\times\left.K_1\left(\frac{\lambda\varrho^2}{4}\right)\right]
-\frac{\sqrt{\pi\lambda}}{\varrho^2}\,
U\left(\frac{3}{2},0,\frac{\lambda\varrho^2}{2}\right),\nonumber
\end{eqnarray}
with $K_n(z)$ being the $n$th-order modified Bessel function of
the second kind and $U(a,b,z)$ being the confluent hypergeometric
function. Fig.~\ref{vdip2dz} shows the function
$[f^z_\lambda(\varrho)]^{-1/3}$ for various values of the trap
aspect ratio $\lambda$. It can be seen that, as expected, the
effective dipolar interaction is repulsive and isotropic.
Independent of $\lambda$, $f^z_\lambda(\varrho)$ asymptotically
approaches $\sqrt{2\pi}\,\varrho^{-3}$ at large $\varrho$. At the
other limit, as $\varrho \rightarrow 0$, $f^z_\lambda(\varrho)$
diverges much slower than $\varrho^{-3}$, although the detailed
behavior at small $\varrho$ depends on the value of $\lambda$.
Note that for a true 2D system with dipoles located in the $xy$
plane and polarized along $z$ axis, one expects a dipolar
interaction potential $\tilde{V}_{dd}^{2D} \propto \varrho^{-3}$
as can be easily seen from Eq.~(\ref{vdd3d}). For the quasi-2D
case considered here, the effective dipolar interaction potential
deviates from $\tilde{V}_{dd}^{2D}$ at small values of $\varrho$
in a fashion that makes the singularity of $V_{dd}^{2D}$ at
$\varrho=0$ integrable, a quite important property for numerical
calculations. An alternative, and often more efficient, way to
treat the dipolar terms in numerical calculations is to use the
Fourier transform of $V_{dd}^{2D}$ \cite{fischer}.

The ground state wave function can be obtained by evolving
Eq.~(\ref{gp2d}) in imaginary time. In the numerical results
presented in the paper, we use $N=10^5$ and $\omega_\perp=2\pi
\times 100$Hz. First we want to study the structure of a single
axial vortex at the center of the cloud. Due to the azimuthal
symmetry, such a state can be written as
$\psi({\boldsymbol\rho})=\chi(\rho)e^{i \varphi}$ with $\varphi$
being the azimuthal angle and $\chi$ a real function satisfying
the following 1D radial equation:
\begin{eqnarray}
i\frac{\partial \chi}{\partial
t}=\left[-\frac{1}{2\rho}\frac{\partial}{\partial\rho}
\left(\rho\frac{\partial}{\partial\rho}\right)
+\frac{1}{2\rho^2}+\frac{\rho^2}{2}+g_{2D}\chi^2+{\cal
D}(\rho)\right]\chi,\nonumber
\end{eqnarray}
where \[ {\cal D}(\rho)\!=\!\!\int_0^\infty \!\!\!\rho'
d\rho'\!\!\int_0^{2\pi}\!\!\!d\varphi'V_{dd}^{2D}(\rho^2+\rho'^2
-2\rho\rho'\cos\varphi')\chi^2(\rho'). \]

\begin{figure}
\centering
\includegraphics[width=2.6in]{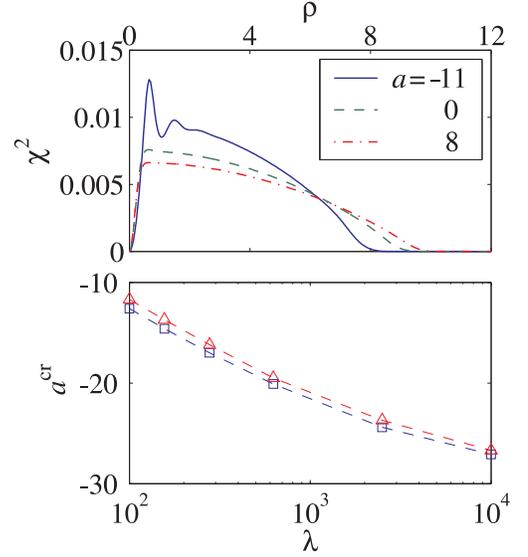}
\caption{(Color online). (a) Radial density profile of the single
vortex state for $\lambda=100$ and various scattering lengths (in
units of Bohr radius, $a_B$). (b) The $\lambda$ dependence of
critical scattering length (in units of $a_B$). Here $\square$ and
$\vartriangle$ denote, respectively, the non-vortex state and the
axial vortex state.} \label{w1dzcri}
\end{figure}

Figure~\ref{w1dzcri}(a) illustrates the radial density profile
$\chi^2(\rho)$ of the single vortex state for different values of
the scattering length. An interesting feature one can notice is
that the density close to the vortex core oscillates as long as
$a$ is not too large. Such oscillatory behavior is induced by the
dipolar interaction and gives the vortex a crater-like shape.
Similar density oscillations are also observed in numerical
studies with other forms of non-local interaction potentials which
are originally employed to model the inter-particle interactions
in superfluid $^4$He \cite{sadd,berloff}. Note that no
oscillations are found in the ground state structures of a
non-rotating dipolar condensate.

The density oscillations of the vortex state also affect its
stability property. A condensate with sufficiently large and
negative $a$ is unstable and tends to collapse \cite{collapse}.
The critical scattering length $a^{\rm cr}$ as a function of the
trap aspect ratio is shown in Fig.~\ref{w1dzcri}(b) for both the
axial vortex state and the non-vortex state. Near $a^{\rm cr}$,
due to the dipole-induced density oscillation, the vortex state
has a peak density exceeds that of the non-vortex state. Since
collapse starts locally at the high density region, the vortex
state tends to be less stable, having a critical scattering length
smaller in magnitude by about 5$\sim$10\% compared with the
non-vortex state. This is in contrast with the non-dipolar
condensate where the vortex states are shown to be more stable
\cite{saito} as they have lower peak densities compared to the
non-vortex states. We remark that the repulsive dipolar
interaction in this case helps to stabilize a condensate with
attractive scattering length: with all the other parameters being
the same as in Fig.~\ref{w1dzcri}(b), a non-dipolar condensate
would have a critical scattering length on the order of $-0.1a_B$.

\begin{figure}
\centering
\includegraphics[width=2.5in]{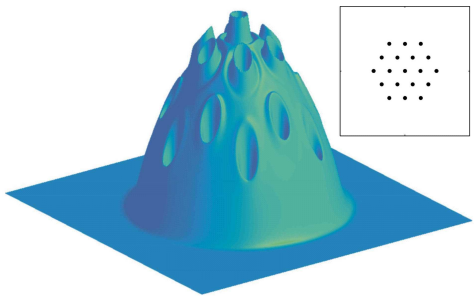}
\caption{(Color online). Density profile of a vortex lattice state
for $\lambda=100$, $\Omega=0.4\omega_\perp$, and $a=-10a_B$. The
inset shows the position of the vortices which form a triangular
lattice with hexagonal symmetry.} \label{crater}
\end{figure}
As we increase the rotation frequency, more vortices enter the
condensate and form a regular lattice as a result of the repulsive
vortex-vortex interaction. For a non-dipolar condensate, it is
well known that the vortex lattice takes a triangular shape with
hexagonal symmetry. Very recently, it is predicted by several
authors that the vortex lattice of a dipolar condensate in the
high-rotating quantum Hall limit may take different geometries
\cite{cooper,zhai}. However, we find from our numerical
calculations that the triangular lattice with hexagonal symmetry
always has the lowest energy for rotating frequencies up to
$\Omega = 0.99 \omega_\perp$. Although computing power and
numerical accuracy currently prevent us from taking higher values
of $\Omega$, this suggests that the quantum Hall regime of a fast
rotating dipolar condensate may need a more careful analysis. An
example of the vortex lattice is shown in Fig.~\ref{crater} where
the crater structure is still present. To ensure that such a
structure is not an artefact due to the reduction of the effective
spatial dimension, we also performed full 3D calculations and
found that the results are in complete agreement with the 2D
calculation.

%\begin{figure}
%\centering
%\includegraphics[width=2.8in]{vdip2dx.eps}
%\caption{(Color online) Function of
%$f_\lambda^x({\boldsymbol\varrho})$, representing the effective 2D
%dipolar interaction potential when the dipoles are polarized along
%the $x$ axis.} \label{vdip2dx}
%\end{figure}

{\em Transversely polarized dipoles} --- It is also of interest to
study the vortex state with dipole moments polarized transversely.
We assume that the dipoles are polarized by a transverse magnetic
field co-rotating with the condensate about the $z$ axis
\cite{note}. Without loss of generality, we assume that the
dipoles are polarized along the $x$ axis in the rotating frame.
The effective quasi-2D dipolar interaction potential for this case
has a similar form as Eq.~(\ref{dip2d}) with
$f^z_\lambda(\varrho)$ replaced by
\begin{eqnarray}
f^x_\lambda({\boldsymbol\varrho})\!&=&\!
\frac{\lambda^{3/2}e^{\lambda\varrho^2/4}}{6}\frac{y^2-2x^2}{\varrho^2}
\left[\lambda\varrho^2K_0
\left(\frac{\lambda\varrho^2}{4}\right)\right.\nonumber\\
&&\!\!\!\!\!\!+\!\left.\left(2-\lambda\varrho^2\right)K_1
\left(\frac{\lambda\varrho^2}{4}\right)\right]
+\frac{\sqrt{\pi\lambda}}{2\varrho^2}\,
U\left(\frac{3}{2},0,\frac{\lambda\varrho^2}{2}\right).\nonumber
\end{eqnarray}
% A plot of $f^x_\lambda$ is shown in Fig.~\ref{vdip2dx} where we
It can be easily shown that, unlike in the previous case, now the
effective 2D dipolar interaction potential is anisotropic: it is
repulsive along $y$ axis and attractive along $x$ axis. Such a
system is unstable for scattering lengths below a threshold value,
which is about $15a_B$ for the parameters used in our calculation.

\begin{figure}
\centering
\includegraphics[width=3.in]{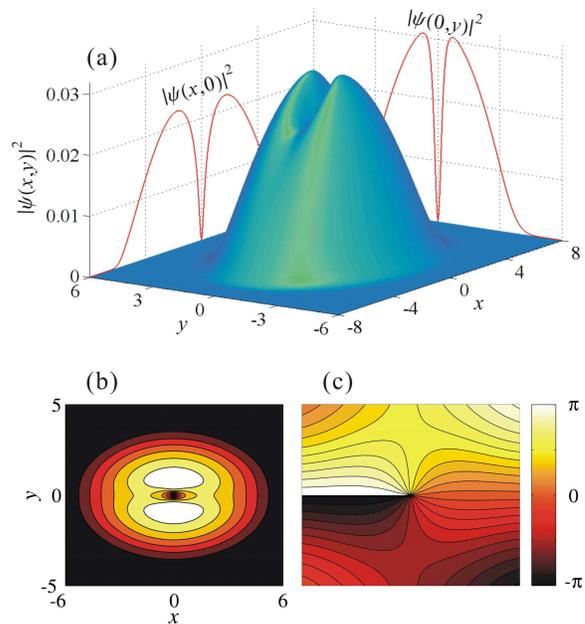}
\caption{(Color online). Structure of a single vortex state with
dipoles polarized along the $x$ axis. The parameters are
$\lambda=100$, $a=16a_B$ and $\Omega=0.3 \omega_\perp$. (a)
Density profile. The solid curves show the densities along the
$x$- and $y$-axis. (b) Density contour plot where the elliptical
shaped vortex core can be clearly seen. Brighter color represents
higher density. (c) Contour plot of the phase of the wave
function.} \label{dipx_single}
\end{figure}

Figure~\ref{dipx_single} shows the structure of a single vortex
state for $a=16a_B$. The whole atomic cloud in this case is
elongated along the $x$ axis with a two-fold symmetry. This is due
to the magnetostriction induced by the dipolar interaction, an
effect that has been recently observed in experiment \cite{mags}.
One can also notice from Fig.~\ref{dipx_single}(b) that the vortex
core is also anisotropic: it has an elliptical shape with the
major axis along $x$. This can be understood as follows: the
vortex core size is determined by the healing length, and the
attractive (repulsive) dipolar interaction along $x$ ($y$)
direction weakens (enhances) the contact interaction, resulting in
a larger (smaller) effective healing length, and hence core size,
along $x$ ($y$).

Along any closed curves around the anisotropic vortex core, there
is still a phase slip of $2 \pi$. However, the phase of the wave
function no longer coincides with the azimuthal angle $\varphi$
[see Fig.~\ref{dipx_single}(c)]. Generally, a vortex state with a
two-fold symmetry can be represented as $\psi \sim \sum_n a_n
\rho^{|2n+1|} e^{i(2n+1)\varphi}$. To a good approximation the
vortex core structure shown in Fig.~\ref{dipx_single} can be
modelled by the three most dominant terms with $n=0$, $\pm 1$,
i.e., the wave function near the core takes the form
\[\psi(\boldsymbol{\rho}) \sim \rho\left(e^{i\varphi}+\alpha
\,e^{-i\varphi}\right)+\beta \rho^3\,e^{3i\varphi} ,\] with
parameters $\alpha<0$ and $\beta>0$.

\begin{figure}
\centering
\includegraphics[width=2.5in]{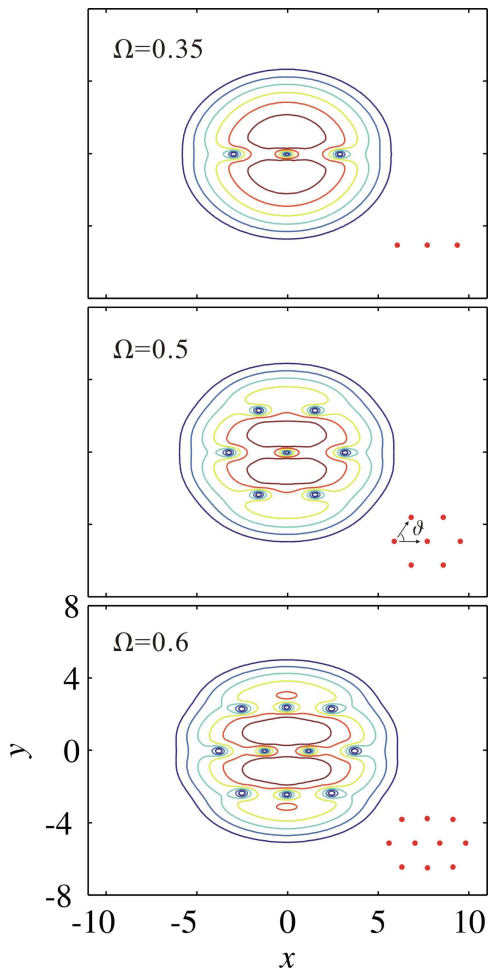}
\caption{(Color online). Density contour plots for $\lambda=100$,
$a=18a_B$ and various rotation frequencies. The dots at the lower
right corner shows the position of vortices. For $\Omega=0.5$ and
$0.6\omega_\perp$, the angles $\vartheta$ are $55^\circ$ and
$62^\circ$, respectively.} \label{dipxlat}
\end{figure}
At higher rotation frequencies, vortex lattice will form with a
structure strongly depending on the scattering length and the
rotating frequency. The richness of the vortex lattice structures
are illustrated in Fig.~\ref{dipxlat} for $a=18a_B$ at three
different rotation frequencies. For $\Omega=0.35\omega_\perp$, three
vortices appear, all located along the $x$ axis. Note that for the
isotropic interaction as in the previous case or in a non-dipolar
condensate, the three-vortex state would have the three vortices
situated on the corners of an equilateral triangle. At higher
rotation frequency, the vortices form lattices with two-fold
symmetry instead of the usual hexagonal symmetry. For $\Omega=0.5$
and 0.6$\omega_\perp$, the angles $\vartheta$ as defined in the
middle plot of Fig.~\ref{dipxlat} are about $55^\circ$ and
$62^\circ$, respectively. Numerically, it becomes increasingly
difficult to find the lowest energy lattice structure when we
further increase $\Omega$. We plan to study this problem in greater
detail in a future work.

We remark that anisotropic vortices have been extensively studied
in the context of high $T_c$ superconductors, where the anisotropy
can arise from the crystal anisotropy \cite{chain} or from high
angular momentum pairing interaction (e.g., in $d$-wave
superconductors) \cite{machida}. In these anisotropic
superconductors, the vortex lattice also deviates from the
hexagonal geometry and can sometimes take a square shape
\cite{gilardi}.

In conclusion, we have investigated the properties of the vortex
states of a quasi-2D rotating dipolar condensate. The vortex
states possess many novel and unique features due to the presence
of the dipolar interaction. We hope that our work will stimulate
more experimental efforts on rotating dipolar quantum gases.

This work is supported by Rice University and the Oak Ridge
Associated Universities.


\begin{thebibliography}{}
\bibitem{note} If the dipoles are not co-rotating with the
condensate, the dipolar interaction potential will have an
explicit time dependence in the rotating frame. Such a system may
not support a stationary state.
\end{thebibliography}
\end{document}